# Imaging currents in HgTe quantum wells in the quantum spin Hall regime

Katja C. Nowack[2,3,*], Eric M. Spanton[1,3], Matthias Baenninger[1,3], Markus König[1,3], John R. Kirtley[2], Beena Kalisky[2,4], C. Ames[5], Philipp Leubner[5], Christoph Brüne[5], Hartmut Buhmann[5], Laurens W. Molenkamp[5], David Goldhaber-Gordon[1,3], Kathryn A. Moler[1,2,3]

[1]Department of Physics, Stanford University, Stanford, California 94305, USA

[2]Department of Applied Physics, Stanford University, Stanford, California 94305, USA

[3]Stanford Institute for Materials and Energy Science, Stanford University, Stanford, California 94305, USA

[4]Department of Physics, Nano-magnetism Research Center, Institute of Nanotechnology and Advanced Materials, Bar-Ilan University, Ramat-Gan 52900, Israel.

[5]Physikalisches Institut (EP3), Universität Würzburg, Am Hubland, D-97074, Würzburg, Germany

*To whom correspondence should be addressed. Email: knowack@stanford.edu

**The quantum spin Hall (QSH) state is a genuinely new state of matter characterized by a non-trivial topology of its band structure[1-5]. Its key feature is conducting edge channels whose spin polarization has potential for spintronic and quantum information applications. The QSH state was predicted[6] and experimentally demonstrated[7] to exist in HgTe quantum wells. The existence of the edge channels has been inferred from the fact that local and non-local conductance values in sufficiently small devices are close to the quantized values expected for ideal edge channels[7,8] and from signatures of the spin polarization[9]. The robustness of the edge channels in larger devices and the interplay between the edge channels and a conducting bulk are relatively unexplored experimentally, and are difficult to assess via transport measurements. Here we image the current in large Hallbars made from HgTe quantum wells by probing the magnetic field generated by the current using a scanning superconducting quantum interference device (SQUID)[10]. We observe that the current flows along the edge of the device in the QSH regime, and furthermore that an identifiable edge channel exists even in the presence of disorder and considerable bulk conduction as the device is gated or its temperature is raised. Our results represent a versatile method for the characterization of new quantum spin Hall materials systems, and**

confirm both the existence and the robustness of the predicted edge channels.

Like an ordinary insulator, the QSH state has a bulk energy gap, but the QSH state supports within the gap a pair of counter-propagating edge modes [1-5]. The QSH state is realized in HgTe quantum wells thicker than the critical thickness of 6.3 nm, whereas thinner quantum wells are ordinary insulators[6,7]. The edge modes are theoretically protected against backscattering by their orthogonal spin states[11,12], and therefore have a quantized conductance of $e^2/h$, where $e$ is the charge of an electron and $h$ is Planck's constant.

Experimentally, nearly-quantized conductance has been found only in devices with edges of several microns or shorter[7,8,13]. For larger devices, the measured resistances deviate from the values expected for dissipationless edge channels[7,8,13]. The edge channels could be affected by spatially-varying quantum well thickness or doping[8,14], magnetic[15] or non-magnetic impurities[16], and Rashba spin-orbit interaction combined with either electron-electron interactions[17] or phonons[18].

To address the edge channels' robustness over larger distances, as well as their interplay with a conducting bulk, we image the magnetic field produced by the current through large Hall bars with a scanning SQUID[10] (Fig. 1a). Images of the current are obtained by deconvoluting the SQUID images (see below). All images are normalized by the applied current. Although most scanning probes cannot image through the top gates, that are usually used to tune carrier density, our technique can.

Hall bars with lateral dimensions shown in Fig. 1b are fabricated from HgTe/(Hg,Cd)Te quantum well structures with nominal well thicknesses (7 nm for H1, 8.5 nm for H2) above the critical thickness[19]. We use only the upper two contacts of the Hall bars to avoid the SQUID touching wire bonds. The two-terminal resistance $R_{2T}$ of H1 is shown in Fig. 1c. We also measured a Hall bar with a quantum well thinner than the critical thickness (Supplementary Information). Measurements are done at $T$ ~3 K unless noted differently.

Fig. 1d and e summarize our main result. When the transport is dominated by bulk conduction (Fig. 1d), the magnetic profile crosses smoothly through zero in the Hall bar, corresponding to homogeneous current flow through the Hall bar. In clear contrast, when the transport is dominated by edge channels (Fig. 1e), the magnetic profile displays two steep crossings through zero at the top and bottom edge of the Hall bar, showing that the current predominantly flows along the edge of the device. The width of features is limited by our spatial resolution (see below). The current along the lower edge fully traces out the perimeter of the top gated part of the Hall bar.

To directly visualize the current density in the device (Fig. 1 f-i), we process the magnetic profiles, which are linked to current density by the Biot-Savart law and a convolution with the SQUID pickup loop. For two-dimensional current flow the Biot-Savart law can be inverted when one magnetic field component is known. We implement the inversion using Fourier transforms[20] and the geometry of the pickup loop as deduced from images of isolated vortices in a bulk superconductor, which act as near-ideal monopole field sources (see Supplementary Information). The current images shown in Fig. 1g and i directly confirm the existence of edge channels in the QSH regime.

The width of the features in the obtained current densities is limited by the 3 μm diameter pickup loop and its scan height of 1.5-2 μm. Systematic errors in the inversion such as ringing and finite current density outside the boundary of the Hall bar result from uncertainty in scan height, imperfect characterization of the pickup loop, noise, and the finite image size (Supplementary Information).

Having compared the extreme cases of bulk- and edge-dominated transport at low and maximum $R_{2T}$, respectively, we next explore the interplay between them. The top gate voltage, $V_{TG}$, tunes the Fermi level from the valence band through the bulk energy gap into the conduction band and thereby changes the bulk conductance, with the bulk being insulating when the Fermi level is in the gap. For a range of $V_{TG}$, we find that edge conduction coexists with bulk conduction (Fig. 2 a-d). In Fig. 2e we show the percentage of current flowing along the edges and bulk obtained by modelling each current profile as a sum of three contributions (Fig. 2f) determined where either the bulk or the edges clearly dominate. The errors are difficult to evaluate (Supplementary Information), but we can identify the presence of a distinct edge current even when the edge carries an order of magnitude less total current than the bulk. Qualitatively, we find that the current flow changes gradually from edge-dominated to bulk-dominated when moving away from the maximum in $R_{2T}$, with a large region of coexistence of edge channels and bulk conduction.

Comparing the top and bottom edges provides additional information about the nature of the edge transport. The percentage of current along the bottom edge initially increases with decreasing $R_{2T}$, since more current arrives at the bottom edge through the bulk of the device. At maximum $R_{2T}$ H1 showed a ratio close to 1:8 between the bottom and top edge current (Fig. 2e), while this ratio is approximately 1:1 in H2 (see below, Fig. 3c). For ballistic edge channels a ratio 1:5 is expected due to the contacts along the bottom path. However, the maximum $R_{2T}$ ~200 kΩ >> $5/6\ h/e^2$, the quantized value expected in our measurement configuration, reveals that the edge channels are not ballistic over their full length as is always seen to date for devices with edges longer than several microns[7,13]. For diffusive transport in the edge channels a ratio ~ 1:1.6 is expected given by the ratio of total edge length along the bottom and top path. The difference between H1 and H2 and the deviation from diffusive transport is consistent with ballistic segments of edge channels of several microns length that are interrupted by scattering sites[8].

It is illustrative to model the bulk and the edges as three parallel resistors. In this simple picture we calculate effective resistances (Fig. 2g) by dividing $R_{2T}$ by the respective current percentage. The bulk resistance resembles an insulator with steep flanks as a function of $V_{TG}$. The resistance of the top edge is rather flat for the $V_{TG}$ range at which it is lower or comparable to the bulk resistance suggesting that the edge channel is relatively unaffected over a significant range of bulk conduction.

As $V_{TG}$ is tuned away from the resistance maximum, the magnetic profile at the top edge broadens for increasing p-type (Fig. 2a) but not for n-type (Fig. 2b) bulk conduction. Different positions along the Hall bar reveal different broadening (Supplementary Information), we therefore attribute the observed broadening to disorder in the current flow and not to a generic broadening of the edge channel.

To further study the robustness of the edge conduction and its coexistence with a conducting bulk, we test how temperature affects the current distribution in H2 (Fig. 3a). The maximum $R_{2T}$ decreases with increasing temperature (Fig. 3b) and the corresponding magnetic profiles show increased bulk conduction. We extract the percentages of edge and bulk current (Fig. 3c) by directly fitting the magnetic profiles (Supplementary Information) and find effective resistances for the top edge, bottom edge, and bulk as described above. The edge resistance is rather temperature independent while the bulk resistance drops steeply with temperature (Fig. 3d), consistent with thermally activated conduction.

An unintentional but serendipitous electrical shock allowed us to investigate the behaviour that is associated with a more disordered sample. We believe that a different thermal cycle in H1 (Fig. 4a) included disordered charge in the gate dielectric from an electrically shocked top gate. Strikingly, the resistance is now much lower, but still displays a peak in transport that could falsely be interpreted as a closer to ideal edge channel. However, imaging reveals that disorder over tens of microns causes the bulk to remain conductive in parts of the Hall bar at every $V_{TG}$ (Fig. 4b-d), and that each region of the Hall bar shows well-defined edge conduction at some $V_{TG}$.

Possible trivial origins for enhanced edge conduction include doping of the edges during fabrication and related band bending. To address this issue, we imaged the current flow in a Hall bar made from a 5nm wide quantum well and found no signature of enhanced edge conductance (Supplementary Information). The evidence for helical edge channels from transport measurements[7-9], the absence of edge currents in the insulating Hall bar, and the presence of edge current throughout the full resistance peak in the non-insulating Hall bars all indicate that our observations originate from the QSH effect.

In conclusion, we directly demonstrate through images the presence of edge channels in HgTe quantum wells in the QSH regime. We observe that edge channels dominate the transport even when the device edges are much longer than the scattering length and that the edge channels persist in the presence of induced disorder and bulk conduction. These properties of the edge channels would be challenging to establish with global transport measurements alone. Our technique is non-invasive and compatible with the use of a top gate, opening the way to explore other topologically non-trivial materials[21-24] with a combination of global transport and local imaging.

**Methods Summary**

Hall bars H1 and H2 were fabricated from two different HgTe/(Hg,Cd)Te quantum well structures with nominal well thickness of 7.0 nm and 8.5 nm (see Supplementary Information). The devices were patterned using optical lithography and subsequent Ar ion-beam etching[19]. The top gates were also patterned using optical lithography and consist of a 40 nm-thick $Al_2O_3$ gate insulator and a Ti (5 nm)/Au (50 nm) gate electrode (see Supplementary Information for additional transport characterization).

All presented images were taken by applying an AC current to the Hall bars and recording the SQUID signal using a lock-in amplifier. The SQUID signal is directly proportional to the flux in the pickup loop caused by the *z*-component of the magnetic field generated by the current. The flux is in units of the magnetic flux quantum, $\Phi_0$. The magnetic images were taken at an rms current amplitude ranging from 100 nA to 500 nA (apart from the profile at $V_{TG}$ = 0.1V in Fig. 2, which was taken at 1 μA). This corresponds to bias voltages that are higher than typically applied in transport experiments. We have explicitly checked that all shown measurements were recorded in or close to the linear regime (Supplementary Information).

**Note** The Goldhaber-Gordon group in collaboration with the Molenkamp group at University of Würzburg have carried out scanning gate measurements on HgTe quantum wells (arXiv:1211.3917, authors: Markus König, Matthias Baenninger, Andrei G. F. Garcia, Nahid Harjee, Beth L. Pruitt, C. Ames, Philipp Leubner, Christoph Brüne, Hartmut Buhmann, Laurens W. Molenkamp, David Goldhaber-Gordon). Our colleagues in the Zhi-Xun Shen group at Stanford, in collaboration with some of us (the Molenkamp group at University of Würzburg and the Goldhaber-Gordon group at Stanford), have obtained images of a HgTe quantum well using microwave impedance microscopy (authors: Yue Ma, Worasom Kundhikanjana, Jing Wang, Reyes Calvo, Biao Lian, Yongliang Yang, Keji Lai, Matthias Baenninger, Markus König, Christopher Ames, Christoph Brüne, Hartmut Buhmann, Philip Leubner, Qiaochu Tang, Kun Zhang, Xinxin Li, Laurens Molenkamp, Shoucheng Zhang, David Goldhaber-Gordon, Michael K. Kelly, Zhi-Xun Shen).

**Supplementary Information** is available at
http://www.stanford.edu/group/moler/publications.html.

**Acknowledgements** We thank S. C. Zhang, X. L. Qi, M. R. Calvo for valuable discussions, H. Noad for assistance with the experiment, G. Stewart for rendering Fig. 1a and M. E. Huber for assistance in SQUID design and fabrication. This work was funded by the Department of Energy, Office of Basic Energy Sciences, Division of Materials Sciences and Engineering, under contract DE-AC02- 76SF00515 (Sample fabrication and scanning SQUID imaging of the QSH state in HgTe Hall bars), by the DARPA Meso project under grant no. N66001-11-1-4105 (MBE growth of the HgTe heterostructures) and by the Center for Probing the Nanoscale, an NSF NSEC, supported under grant no. PHY-0830228 (development of the scanning SQUID technique). The work at Würzburg was also supported by the German research foundation DFG (SPP 1285 'Halbleiter Spintronik' and DFG-JST joint research program 'Topological Electronics') and by the EU through the ERC-AG program (project '3-TOP'). B.K. acknowledges support from FENA.

**Author Contributions** K. C. N. and E. M. S. performed the SQUID measurements. K. C. N., E. M. S., B. K., J. R. K analysed the results with input from K. A. M., D. G. G., M. K. , M. B.. M. B. fabricated the samples. C. A., P. L., C. B., H. B. and L. W. M. grew the quantum well structures. K. A. M., D. G. G. and L. W. M. guided the work. K. C. N. and K. A. M. wrote the manuscript with input from all co-authors.

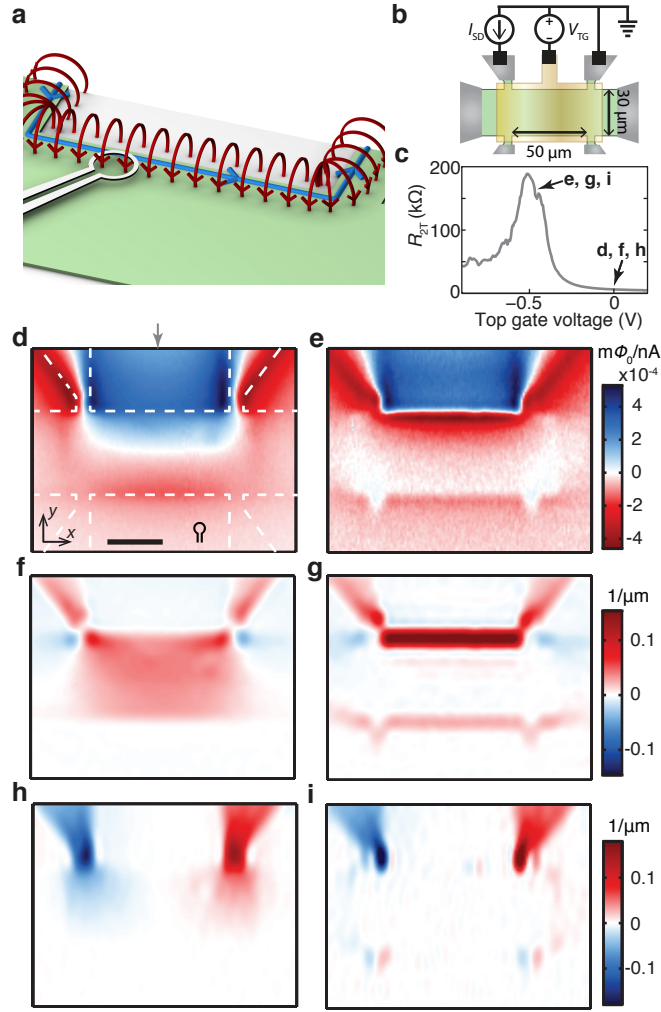

**Figure 1. Current flows along the edge in the QSH regime. a**, Sketch of the measurement. The magnetic field (red) generated by the current (blue) is measured by detecting the flux through the SQUID's pickup loop. **b**, Schematic of the Hall bar. **c**, Two terminal resistance $R_{2T}$ of H1 vs. top gate voltage $V_{TG}$. **d, e,** Flux images at $V_{TG}$ as indicated in **c** measured on H1. In **d** a 20 μm scalebar (black), the outline of the Hall bar mesa (white dashed line) and a sketch of the pickup loop (black) are included; grey arrow indicates the x-position of the profiles in Fig. 2. **f, g,** $X$-component and **h,i,** $y$-component of the two-dimensional current density obtained from the current inversion of the flux images in **d** and **e** respectively.

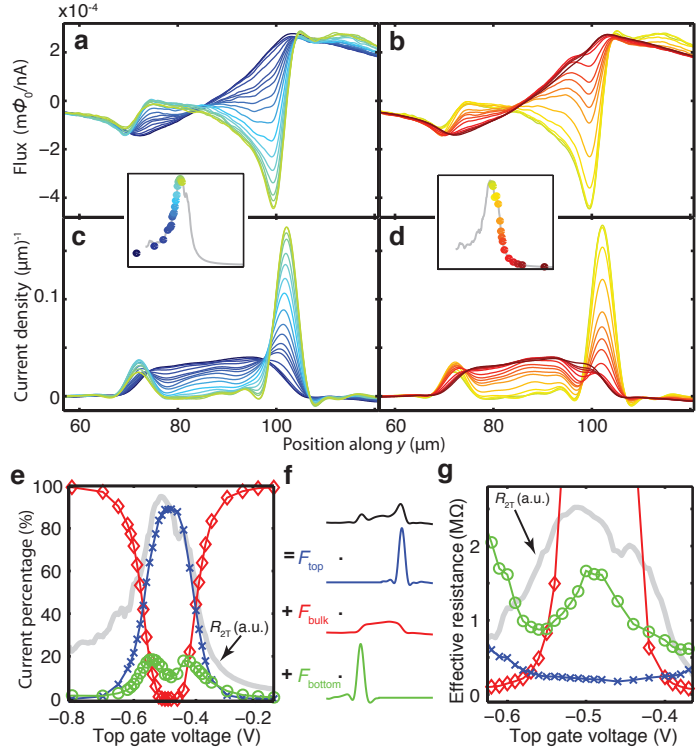

**Figure 2. Coexistence of edge channels and a conductive bulk. a, b,** Flux profiles along the y-direction at the position as indicated in Fig. 1d as a function of $V_{TG}$. Insets: $R_{2T}$ from Fig. 1b, dot colours match the profile colours to indicate $V_{TG}$. **c, d,** Current profiles at the same position and with the same colour coding as in **a, b**. All line cuts are averaged over a width of ~ 5 μm. Integration of the current profiles given in rescaled units μm$^{-1}$ yields 1.0 +/- 0.05, as expected.
**e,** Percentage of current flowing along the top edge (blue crosses), the bottom edge (green circles) and through the bulk (red diamonds) obtained through modelling each current profile in **c, d** by a sum of a bulk and two edge contributions as sketched in **f**, where amplitudes $F_{top}$, $F_{bulk}$ and $F_{bottom}$ give the current percentage. **g,** Effective resistances of the bulk (symbols and colours as in **e**) and the edges obtained from dividing the two-terminal resistance $R_{2T}$ by the current fractions from **e** at each $V_{TG}$. $V_{TG}$ is restricted to values at which $F_{top} > 10\%$. Grey lines in **e** and **g** are $R_{2T}$ from Fig. 1c in a.u.

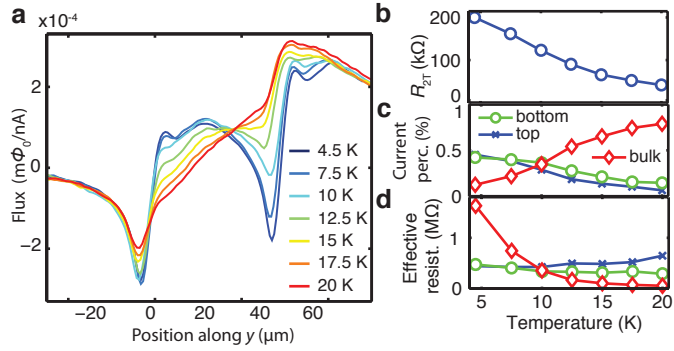

**Figure 3. Temperature dependence. a,** Flux profiles as a function of temperature measured on Hall bar H2. $V_{TG}$ is adjusted for each profile, such that $R_{2T}$ is at its maximum. **b,** Maximum value of $R_{2T}$ as a function of temperature. **c,** Percentage of current flowing along the top and bottom edge and through the bulk, extracted from fitting the flux profiles in a with a bulk and two edge contributions. **d,** Effective resistance of the bulk and the edges obtained from dividing $R_{2T}$ by the current percentage.

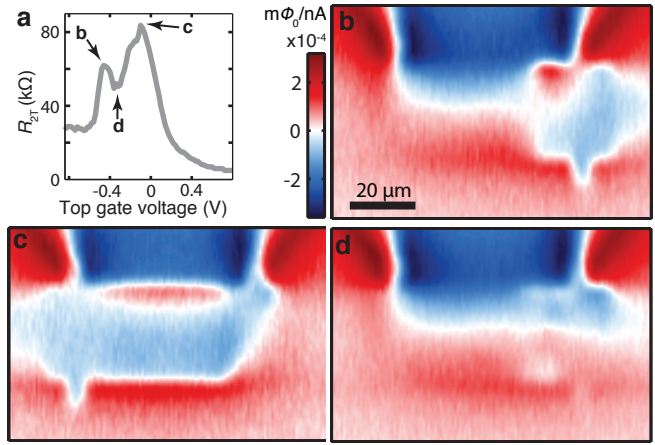

**Figure 4. Gate dielectric induced disorder. a,** Two terminal resistance of H1 (same device as in Fig. 1) but measured in a different thermal cycle. **b-d,** Flux images at top gate voltages as indicated in **a**.